\documentstyle[12pt]{article}
\begin{document}
\renewcommand{\theequation}{\thesection.\arabic{equation}}
\newcommand{\eqn}[1]{(\ref{#1})}
\renewcommand{\section}[1]{\addtocounter{section}{1}
\vspace{5mm} \par \noindent
  {\bf \thesection . #1}\setcounter{subsection}{0}
  \par
   \vspace{2mm} } 
\newcommand{\sectionsub}[1]{\addtocounter{section}{1}
\vspace{5mm} \par \noindent
  {\bf \thesection . #1}\setcounter{subsection}{0}\par}
\renewcommand{\subsection}[1]{\addtocounter{subsection}{1}
\vspace{2.5mm}\par\noindent {\em \thesubsection . #1}\par
 \vspace{0.5mm} }
\renewcommand{\thebibliography}[1]{ {\vspace{5mm}\par \noindent{\bf
References}\par \vspace{2mm}}
\list
 {\arabic{enumi}.}{\settowidth\labelwidth{[#1]}\leftmargin\labelwidth
 \advance\leftmargin\labelsep\addtolength{\topsep}{-4em}
 \usecounter{enumi}}
 \def\newblock{\hskip .11em plus .33em minus .07em}
 \sloppy\clubpenalty4000\widowpenalty4000
 \sfcode`\.=1000\relax \setlength{\itemsep}{-0.4em} }
\def\bea{\begin{eqnarray}}
\def\eea{\end{eqnarray}}
\def\be{\begin{equation}}
\def\ee{\end{equation}}
\def\alp{\alpha}
\def\bet{\beta}
\def\gam{\gamma}
\def\del{\delta}
\def\eps{\epsilon}
\def\sig{\sigma}
\def\lam{\lambda}
\def\Lam{\Lambda}
\def\ss{\scriptstyle}
\newcommand\rf[1]{(\ref{#1})}
\def\nn{\nonumber}
\newcommand{\sect}[1]{\setcounter{equation}{0} \section{#1}}
\renewcommand{\theequation}{\thesection .\arabic{equation}}
\newcommand{\NPB}[3]{{Nucl.\ Phys.} {\bf B#1} (#2) #3}
\newcommand{\CMP}[3]{{Commun.\ Math.\ Phys.} {\bf #1} (#2) #3}
\newcommand{\PRD}[3]{{Phys.\ Rev.} {\bf D#1} (#2) #3}
\newcommand{\PLB}[3]{{Phys.\ Lett.} {\bf B#1} (#2) #3}
\newcommand{\JHEP}[3]{{JHEP} {\bf #1} (#2) #3}
\newcommand{\ft}[2]{{\textstyle\frac{#1}{#2}}}
\def\st{\scriptstyle}
\def\sst{\scriptscriptstyle}
\def\mco{\multicolumn}
\def\epp{\epsilon^{\prime}}
\def\vep{\varepsilon}
\def\ra{\rightarrow}
\def\al{\alpha}
\def\ab{\bar{\alpha}}
\renewcommand{\a}{\alpha}
\renewcommand{\b}{\beta}
\renewcommand{\c}{\gamma}
\renewcommand{\d}{\delta}
\newcommand{\pa}{\partial}
\newcommand{\g}{\gamma} \newcommand{\G}{\Gamma}
\newcommand{\e}{\epsilon}
\newcommand{\z}{\zeta}
\renewcommand{\l}{\lambda}
\newcommand{\m}{\mu}
\newcommand{\n}{\nu}
\newcommand{\x}{\chi}
\newcommand{\p}{\pi}
\newcommand{\s}{\sigma}
\renewcommand{\t}{\tau}
\newcommand{\y}{\upsilon}
\renewcommand{\o}{\omega}
\newcommand{\q}{\theta}
\newcommand{\h}{\eta}
\newcommand\text[1]{\rm #1}
\newcommand\WL{{dWL}}
\newcommand\order[1]{\vert_{\theta^#1}}
\newcommand\Border[1]{\Big\vert_{\theta^#1}}
\newcommand\km{\kappa_-}

\thispagestyle{empty}

\begin{center}

\hfill THU-98/27\\
\hfill NIKHEF 98-024\\
\hfill VUB/TENA/98/4\\[3mm]
\hfill{\tt hep-th/9808052}\\
 
\vspace{3cm}

{\large\bf The M-Theory Two-Brane in AdS$_4\times $S$^7$ 
and AdS$_7\times $S$^4$}\\
\vspace{1.4cm}
{\sc Bernard de Wit ${}^a$, Kasper Peeters $^b$,
Jan Plefka ${}^b$ \\and Alexander Sevrin ${}^c$ }\\

\vspace{1.3cm}

${}^a${\em Institute for Theoretical Physics, Utrecht University}\\
{\em Princetonplein 5, 3508 TA Utrecht, The Netherlands}\\
{\footnotesize \tt bdewit@phys.uu.nl}\\

\vspace{.5cm}

${}^b${\em NIKHEF, P.O. Box 41882, 1009 DB Amsterdam,}\\
{\em The Netherlands}\\
{\footnotesize \tt t16@nikhef.nl, plefka@nikhef.nl}\\

\vspace{.5cm}

${}^c${\em Theoretische Natuurkunde, Vrije Universiteit Brussel} \\
{\em Pleinlaan 2, B-1050 Brussel, Belgium} \\
{\footnotesize \tt asevrin@tena4.vub.ac.be}\\
\end{center}

\vspace{1.2cm}

\centerline{\bf Abstract}
\vspace{- 4 mm}  
\begin{quote}\small
We construct the supermembrane
action in an  $AdS_4\times S^7$ and $AdS_7\times S^4$ background to all 
orders in anticommuting coordinates. The result is compared to and 
agrees completely with results obtained earlier for generic supergravity
backgrounds through gauge completion
at low orders in $\theta$.
\end{quote}
\vfill
\leftline{\sc August 1998}
\newpage
\setcounter{page}{1}
\baselineskip18pt

\addtocounter{section}{1}
\par \noindent
{\bf \thesection . Introduction}
  \par
   \vspace{2mm} 
\noindent
The two-brane (as well as its dual five-brane) plays a central 
role in M-theory \cite{townsend}. Recently the interactions of 
this supermembrane   
\cite{BST} in a background described by the component fields of 
11-dimensional supergravity \cite{CJS} were written down in 
low orders of the anticommuting coordinates $\theta$ 
\cite{backgr}. In principle this offers the possibility for 
studying the two-brane in a number of interesting supergravity 
backgrounds with a high degree of supersymmetry. In view of 
the recent connection noted between the near-horizon D-brane 
solutions and certain superconformal field theories 
\cite{maldacena}, it is of interest to study the supermembrane in 
backgrounds of an anti-de Sitter spacetime times a compact 
manifold. This program was recently carried out for the 
type-IIB superstring and the D3-brane
in a IIB-supergravity background of this type 
\cite{MT,KRR,MT2}. In the context of 11-dimensional 
supergravity the $AdS_4\times S^7$ and $AdS_7\times S^4$ 
backgrounds stand out as they leave 32 supersymmetries 
invariant \cite{sevenS,fourS}. These backgrounds  
are associated with the near-horizon geometries corresponding to 
two- and five-brane configurations and thus to possible 
conformal field theories in 3 and 6 spacetime dimensions with 16 
supersymmetries, whose exact nature is not yet completely known. 

The purpose of this paper is to discuss the supermembrane in 
these two backgrounds.  As these spaces are local products of
homogeneous spaces, their geometric information can be extracted 
from appropriate coset representatives leading to standard 
invariant one-forms corresponding to the vielbeine and 
spin-connections. Our approach differs from that of a recent 
paper \cite{DFFFTT} where one constructs the geometric information 
exploiting simultaneously the kappa symmetry of the supermembrane 
action, in that we determine the geometric information 
independent from the supermembrane action. As this construction 
holds to all 
orders in the fermionic coordinates $\theta$, it provides 
valuable complementary information to the low-order $\theta$ 
results obtained by gauge completion \cite{backgr}. 

The antisymmetric four-rank field strength of M-theory induces the  
compactifications of the theory to $AdS_4\times S^7$ and 
$AdS_7\times S^4$, which leave the 32 supersymmetries intact. 
These two compactifications are thus governed by the 
Freund-Rubin field $f$, defined by (in Pauli-K\"all\'en 
convention, so that we  
can leave the precise signature of the spacetime open),  
\be
F_{\mu\nu\rho\sigma}=6 f \,e\, \varepsilon_{\mu\nu\rho\sigma}\,, 
\label{freundrubin}
\ee
with $e$ the vierbein determinant. 
When $f$ is purely imaginary we are dealing with an $AdS_4\times 
S^7$ background while for real $f$ we have an $AdS_7\times S^4$ 
background. The nonvanishing curvature components corresponding to
the 4- and 7-dimensional subspaces are equal to 
\bea
R_{\mu\nu\rho\sigma}&=&- 4 f^2 ( g_{\mu\rho}\,g_{\nu\sigma}
-g_{\mu\sigma}\,g_{\nu\rho})\,,
\nn\\
R_{\mu'\nu'\rho'\sigma'}&=& f^2 ( g_{\mu'\rho'}\,g_{\nu'\sigma'}
-g_{\mu'\sigma'}\,g_{\nu'\rho'}) \,. \label{curvatures}
\label{sugrasol}
\eea
Here $\m,\n,\rho,\s$ and $\m^\prime,\n^\prime,
\rho^\prime ,\s^\prime$ are 4- and 7-dimensional world indices, 
respectively. We also use $m_{4,7}$ for the inverse radii of the 
two subspaces, defined 
by $\vert f\vert^2={m_7}^2=\ft14{m_4}^2$.
The Killing-spinor equations associated with the 32 
supersymmetries in this background take the form
\bea
\Big(D_\m - f\gamma_\m \gamma_5\otimes {\bf 1} \Big) \epsilon = 
\Big(D_{\m^\prime} +\ft12 f{\bf 1}\otimes\gamma_{\m^\prime}^\prime \Big) 
\epsilon = 0  \,, \label{killing-spinors}
\eea
where we make use of the familiar decomposition of the 
(hermitean) gamma matrices appropriate to the 
product space of a 4- and a 7-dimensional subspace\footnote{%
  The 11-dimensional gamma matrices decompose 
  into the ones referring to the 4-space and the 7-space,  
  denoted by $\gamma^r$ and $\gamma^\prime_{\!r^\prime}$, 
  respectively. Hence the 4- and 7-dimensional tangent-space 
  indices are denoted by 
  $r,s,\ldots$ and $r^\prime, s^\prime,\ldots$; 
  11-dimensional  tangent-space indices are distinguished by a 
  caret. The 11-dimensional gamma matrices then take the form
  $\Gamma_r=\gamma_r\otimes{\bf 1}$ and  
  $\Gamma_{r'}=\gamma_5\otimes\gamma'{}_{\!r'}$. The 
  charge-conjugation matrix decomposes into  
  the ones referring to the 4- and 7-dimensional subspaces 
  according to ${\cal C}=C\otimes C'$.  
  Observe that $C$ is antisymmetric and $C^\prime$ is symmetric, 
  so that $\cal C$ is always antisymmetric. Furthermore we have 
  $C^{-1}\gamma_rC = -\gamma_r{}^{\!\rm T}$ and 
  $C^{\prime -1}\gamma^\prime_{\!r^\prime}C'=-\gamma^\prime_{\!
  r^\prime}{}^{\!\rm T}$ 
  and $\gamma_5=\frac{1}{24}\varepsilon^{rstu} 
  \gamma_r\gamma_s\gamma_t\gamma_u$. For future use we also note 
  that spinorial tangent-space indices are denoted by $a,b,\ldots$, 
  $a^\prime,b^\prime,\ldots$ and $\hat a,\hat b,\ldots$ for 4-, 7- 
  or 11-dimensional indices. In the gauge we are working in there 
  will be no distinction between tangent and world spinor indices.
  Gamma matrices with multiple 
  indices denote weighted antisymmetrized products of gamma 
  matrices in the standard way.
}. %
Here $D_\m$ and $D_{\m^\prime}$ denote the covariant derivatives 
containing the spin-connection fields corresponding to SO(3,1) or 
SO(4) and SO(7) or SO(6,1), respectively.


\setcounter{equation}{0}
\section{Structure of $osp(8|4)$ and $osp(6,2|4)$}
\noindent
The algebra of isometries of the $AdS_4\times S^7$ and 
$AdS_7\times S^4$ backgrounds is given by $osp(8|4)$ and 
$osp(6,2|4)$. Their bosonic subalgebra consists of 
$so(8)\oplus sp(4)\simeq so(8)\oplus so(3,2)$ and $so(6,
2)\oplus usp(4)\simeq so(6,2)\oplus so(5)$, respectively. 
The spinors transform in the $(8,4)$ of this 
algebra. Observe that the spinors transform in a chiral 
representation of $so(8)$ or $so(5)$. 

We decompose the generators of $osp(8|4)$ or $osp(6,2|4)$ in 
terms of irreducible representations of the bosonic $so(7)\oplus 
so(3,1)$ and $so(6,1)\oplus so(4)$ subalgebras. In that way we get 
the bosonic (even)  generators $P_r$, $M_{rs}$, which generate $so(3,
2)$ or $so(5)$,  and $P_{r'}$, $M_{r's'}$, which generate $so(8)$ 
or $so(6,2)$. All the bosonic generators are taken antihermitean 
(in the Pauli-K\"all\'en sense). The fermionic 
(odd) generators $Q_{a a'}$ are Majorana spinors.
The commutation relations between even generators are
\be
\begin{array}{rcl}
{[}P_r,P_s{]}&\!\!=\!\!&-4f^2 M_{rs}\,,\\[2mm]
{[}P_r,M_{st}{]}&\!\!=\!\!&\delta_{rs}\,P_t-\delta_{rt}\,P_s\,,
\\[2mm]
{[}M_{rs},M_{tu}{]}&\!\!=\!\!&\delta_{ru}\,M_{st}+\delta_{st}\,
M_{ru}\\[1mm]
&&-\delta_{rt}\,M_{su}-\delta_{su}\,M_{rt}\,,
\end{array}
\begin{array}{rcl}
{[}P_{r'},P_{s'}{]}&\!\!=\!\!&f^2 M_{r's'}\,,\\[2mm]
{[}P_{r'},M_{s't'}{]} &\!\!=\!\!& \delta_{r's'}\,P_{t'}-\delta_{r't'}\,
P_{s'}\,,\\[2mm]
{[}M_{r's'},M_{t'u'}{]}&\!\!=\!\!&\delta_{r'u'}\,M_{s't'}+\delta_{s't'}\,
M_{r'u'}\\[1mm] 
&& -\delta_{r't'}\,M_{s'u'}-\delta_{s'u'}\,M_{r't'}\,.
\end{array} \label{bosonic-comm}
\ee
The odd-even commutators are given by
\be
\begin{array}{rcl}
{[}P_r,Q_{a a'}{]}&\!\!=\!\!&- {f} (\gamma_r\gamma_5)_a {}^b \,Q_{b 
a'}\,,\\[2mm]
{[}M_{rs},Q_{a a'}{]}&\!\!=\!\!& -\ft12(\gamma_{rs})_a {}^b \,Q_{ba 
'}\,,
\end{array}
\quad
\begin{array}{rcl}
{[}P_{r'},Q_{a a'}{]}&\!\!=\!\!& -\ft12 {f}(\gamma'{}_{\!r'})_{a'} {}^{b'} 
\,Q_{a b'}\,,\\[2mm]
{[}M_{r's'},Q_{a a'}{]}&\!\!=\!\!& -\ft12 (\gamma'{}_{\!r's'})_{a'} {}^{b'} 
\,Q_{a b'}\,.
\end{array}
\ee
Finally, we have the odd-odd anti-commutators, 
\begin{eqnarray}
\{Q_{a a '},Q_{bb'}\}&=&-(\gamma_5 C)_{a b}
\left(2(\gamma'{}_{\!r'}C')_{a 'b'}\,P^{r'}
- f(\gamma'{}_{\!r's'}C')_{a 'b'}\,M^{r's'}\right) \nonumber\\
&&-C'{}_{\!a' b'}\Bigl(2(\gamma{}_{r}C)_{a b}\,P^{r}
+ 2  f (\gamma{}_{rs}\gamma_5 C)_{a b}\,M^{rs}\Bigr ).
\end{eqnarray}
All other (anti)commutators vanish. The normalizations of the
above algebra were determined by comparison with the 
supersymmetry algebra in the conventions of \cite{backgr} 
in the appropriate backgrounds. In fact it is convenient for 
later application to 
recast the above results again in 11-dimensional form.
The (anti)commutation relations that involve the supercharges 
then read as,
\begin{eqnarray}
{[}P_{\hat r},\bar Q{]}&\!\!=\!\!& \bar Q \, T_{\hat r}{}^{\!\hat s 
\hat t\hat u\hat v} F_{\hat s\hat t\hat u\hat v} \,,\qquad
{[}M_{\hat r\hat s},\bar Q{]}= \ft12 \bar Q\,\Gamma_{\hat r\hat s} \,, 
\nonumber\\
\{Q,\bar Q\}&\!\!=\!\!&-2 \Gamma_{\hat r}\,P^{\hat r}
+ \ft1{144} \Big[ \Gamma^{\hat r\hat s\hat t\hat u\hat v\hat w} 
F_{\hat t\hat u\hat v\hat w} 
+ 24 \,\Gamma_{\hat t\hat u}   F^{\hat r\hat s\hat t\hat u} \Big] 
M_{\hat r\hat s} \,, \label{11-comm}
\end{eqnarray}
where $T$ is the following combination of $\Gamma$-matrices,
\be
T_{\hat r}{}^{\!\hat s \hat t\hat u\hat v}= \ft1{288} \Big(
\Gamma_{\hat r}{}^{\!\hat s \hat t\hat u\hat v} - 8 \,\delta_{\hat 
r}{}^{\![\hat s} \,\Gamma^{\hat t\hat u\hat v]}\Big) \,.
\ee
The above formulae are only applicable in the background 
where the field strength takes the form given in 
\eqn{freundrubin}. 


\setcounter{equation}{0}
\section{The coset space representations of $AdS_4\times S^7$ 
and $AdS_7\times S^4$} 
\noindent
Both backgrounds that we consider correspond to homogenous spaces 
and can thus be formulated as coset spaces \cite{cosets}. 
In the case at hand 
these (reductive) coset spaces $G/H$ are $OSp(8|4)/[SO(7)\times SO(3,1)]$ 
and $OSp(6,2|4)/[SO(6,1)\times SO(4)]$. 
To each element of the coset $G/H$ we associate an element of 
$G$, which we denote by $L(Z)$. Here $Z^A$ stands for the 
coset-space coordinates $x^{\hat r}$, $\theta^{\hat a}$ (or, 
alternatively, $x^r$, $y^{r'}$ and $\theta^{a a '}$). The 
coset representative $L$ transforms from the left under constant 
$G$-transformations corresponding to the isometry group of the 
coset space and from the right under local $H$-transformations: 
$L\to L^\prime = g\,L\,h^{-1}$. 

The vielbein and the torsion-free $H$-connection one-forms, $E$ 
and $\Omega$, are defined through\footnote{%
  A one-form $V$ stands for $V\equiv {\rm d}Z^AV_A$ and an exterior 
  derivative acts according to ${\rm d}V\equiv -{\rm d}Z^B\wedge 
  {\rm d}Z^A \, \partial_AV_B$. Fermionic derivatives are thus 
  always left-derivatives.} 
\be
{\rm d} L + L\,\Omega = L\, E\,,\label{defv}
\ee
where 
\begin{eqnarray}
E= E^{\hat r}P_{\hat r} +\bar E Q \,,\qquad 
\Omega= \ft12 \Omega^{\hat r\hat s}M_{\hat r \hat s}.
\end{eqnarray}
The integrability of \eqn{defv} leads to the Maurer-Cartan 
equations, 
\begin{eqnarray}
\label{maurercartan}
&&{\rm d}\Omega - \Omega \wedge \Omega - \ft12  E^{\hat r}\wedge 
E^{\hat s} \, [P_{\hat r},P_{\hat s}]
- \ft1{288} \bar E\Big[ \Gamma^{\hat r\hat s\hat t\hat u\hat v\hat w} 
F_{\hat t\hat u\hat v\hat w} 
+ 24 \,\Gamma_{\hat t\hat u}   F^{\hat r\hat s\hat t\hat u} \Big] 
E\,M_{\hat r\hat s} =0  \,, \nn\\
&&{\rm d} E^{\hat r} -\Omega^{\hat r\hat s}\wedge E_{\hat s} - 
\bar E\,\Gamma^{\hat r} \wedge E=0 \, ,\nn \\
&&{\rm d} E +  E^{\hat r}\wedge T_{\hat r}{}^{\hat t\hat u\hat 
v\hat w}E \,F_{\hat t \hat u \hat v \hat w} - 
\ft14 \Omega^{\hat r\hat s} \wedge \Gamma_{\hat r\hat s} E=0 \, ,
\end{eqnarray}
where we suppressed the spinor indices on the anticommuting 
component $E^{\hat a}$. 
The first equation in a fermion-free background reproduces 
\eqn{curvatures} upon using the commutation relations 
\eqn{bosonic-comm}.   

The purpose of this section is to determine the vielbeine and 
connections to all orders in $\theta$ for the spaces of interest. 
The choice of the coset representative amounts to a gauge choice 
that fixes the parametrization of the coset space. We will 
not insist on an explicit parametrization of the bosonic part of 
the space. It turns out to be advantageous to factorize $L(Z)$ 
into a group element of the bosonic part of $G$ corresponding 
to the bosonic coset space, whose parametrization we leave 
unspecified, and a fermion factor. Hence we write 
\be
L(Z) =  \ell(x)\, \hat L(\theta)\,, \quad \mbox{ with } \hat 
L(\theta) = \exp [\,\bar \theta Q\,]\,. 
\ee
Following \cite{MT,KRR} and \cite{ssp}, we rescale the odd  
coordinates according to $\theta\rightarrow t\, \theta$, where $t$ is 
an auxiliary parameter that we will put to unity at the end. 
Taking the derivative with respect to $t$ of \eqn{defv} then 
leads to a first-order differential equation for $E$ and $\Omega$ (in 
11-dimensional notation),
\begin{eqnarray}
\dot E - \dot \Omega &=& {\rm d}\bar \theta \,Q  + (E-\Omega) \,
\bar\theta Q - \bar\theta Q \,(E-\Omega) 
\end{eqnarray}
After expanding $E$ and $\Omega$ on the right-hand side in 
terms of the generators and using the (anti)commutation relations 
\eqn{11-comm} we find the coupled differential equations, 
\begin{eqnarray}
\dot E^{\hat a} &=& \Bigl({\rm d}\theta +  E^{\hat r} \,
T_{\hat r}{}^{\!\hat s \hat t\hat u\hat v} \theta \,
F_{\hat s\hat t\hat u\hat v}  -  \ft14 \Omega^{\hat r\hat s}\,
\Gamma_{\hat r\hat s}\theta \Bigr)^{\hat a}\,,\nonumber \\ 
\dot E^{\hat r} &=& 2\, \bar \theta \,\Gamma^{\hat r} E\,,\nonumber 
\\
\dot \Omega^{\hat r\hat s} &=&  \ft1{72} \bar \theta \Big[ 
\Gamma^{\hat r\hat s\hat t\hat u\hat v\hat w}  
F_{\hat t\hat u\hat v\hat w} 
+ 24 \,\Gamma_{\hat t\hat u}   F^{\hat r\hat s\hat t\hat u} \Big] 
E\,. \label{diffeq}
\end{eqnarray}

These equations can be solved straightforwardly \cite{KRR}. For 
instance, we can  
determine the corresponding equations for multiple 
$t$-derivatives, after which we explicitly construct the solution by a 
formal Taylor expansion about $t=0$. Of course, this expansion 
will constitute only a finite series, as the $n$-th derivative 
will be proportional to the $n$-th power of $\theta$.   
A crucial role is played by the initial conditions,
\begin{eqnarray}
E^{\hat a}\vert_{t=0} &=& 0\,,  \nonumber \\
E^{\hat r}\vert_{t=0} &=& e^{\hat r}(x)  \equiv  {\rm d}x^{\hat 
\mu}\,e_{\hat \mu}{}^{\!\hat r}(x)\,,  \nonumber \\ 
\Omega^{\hat r\hat s} \vert_{t=0} &=&  \omega^{\hat r \hat 
s}(x)  \equiv {\rm d}x^{\hat \mu}\, \omega_{\hat\mu}{}^{\!\hat r \hat s}(x)   \,,
\end{eqnarray}
where $e_{\hat \mu}{}^{\!\hat r}$ and $\omega_{\hat\mu}{}^{\!\hat 
r \hat s}$ denote the vielbein and spin-connection components of 
the product of the $AdS$ space and the sphere. From the initial  
conditions it follows that the only nonvanishing derivative at 
$t=0$ equals 
\be
D\theta^{\hat a} \equiv \dot E^{\hat a}\Big\vert_{t=0} =  \Big({\rm d}\theta +  e^{\hat r} \,
T_{\hat r}{}^{\!\hat s \hat t\hat u\hat v} \theta  \, 
F_{\hat s\hat t\hat u\hat v}  -  \ft14 \omega^{\hat r\hat s}\,
\Gamma_{\hat r\hat s}\theta\Big)^{\hat a}\,. 
\ee

We can now give the explicit solution (setting $t=1$), 
\bea
E(x,\theta)&=&\sum_{n=0}^{16}\, \frac{1}{(2n+1)!}\, {\cal M}^{2n} 
\, D\theta\,,  \nn \\ 
E^{\hat r}(x,\theta)&=&{\rm d}x^{\hat \mu}\, e_{\hat \mu} 
{}^{\!\hat r}  + 2 \sum_{n=0}^{15}\,\frac{1}{(2n+2)!}\, 
\bar\theta\,\Gamma^{\hat r}\, {\cal M}^{2n}\, D\theta
\\[2.5 mm]
\Omega^{\hat r\hat s}(x,\theta)&=&{\rm d}x^{\hat \mu}\,  
\omega_{\hat \mu}{}^{\!\hat r\hat s}  \nn \\
&& +\ft1{72}  
\sum_{n=0}^{15}\frac{1}{(2n+2)!}\, \bar\theta \,[ 
\Gamma^{\hat r\hat s\hat t\hat u\hat v\hat w}  
F_{\hat t\hat u\hat v\hat w} 
+ 24 \,\Gamma_{\hat t\hat u}   F^{\hat r\hat s\hat t\hat u} ] 
\, {\cal M}^{2n}\, D\theta    \,, 
\nn
\eea
where the matrix ${\cal M}^2$ equals \cite{KRR},
\bea
({\cal M}^2)^{\hat a}{}_{\!\hat b}  &=& 2\, (T_{\hat r}{}^{\!\hat s 
\hat t\hat u\hat v}\, \theta )^{\hat a}\, 
F_{\hat s\hat t\hat u\hat v} \, (\bar \theta \,\Gamma^{\hat 
r})_{\hat b} \nn\\
&& - \ft1{288} (\Gamma_{\hat r\hat s}\, \theta)^{\hat 
a}\, (\bar\theta\,[\Gamma^{\hat r\hat s\hat t\hat u\hat v\hat w}  
F_{\hat t\hat u\hat v\hat w} 
+ 24 \,\Gamma_{\hat t\hat u}   F^{\hat r\hat s\hat t\hat 
u}])_{\hat b}\,.
\eea
The lowest-order terms in these expansions are given by 
\begin{eqnarray}
\label{vielbeinexp}
E^{\hat r} &=& e^{\hat r} + \bar\theta\Gamma^{\hat r} {\rm 
d}\theta +  
\bar\theta \Gamma^{\hat r} ( e^{\hat m}\,T_{\hat m}{}^{\hat s\hat 
t\hat u\hat v}  
  F_{\hat s\hat t\hat u\hat v} - \ft14 \omega^{\hat s\hat t} \,
\Gamma_{\hat s\hat t})\theta 
+ {\cal O}(\theta^4)\, ,\nn \\
E &=& {\rm d}\theta + 
( e^{\hat r}\,T_{\hat r}{}^{\hat s\hat t\hat u\hat v}
F_{\hat s\hat t\hat u \hat v} - \ft14 \omega^{\hat r\hat s}\, 
\Gamma_{\hat r\hat s}) \theta + {\cal O}(\theta^3)   \,,\nn \\
\Omega^{\hat r\hat s} &=& \omega^{\hat r \hat s}  +\ft1{144}  
 \bar\theta \,[ 
\Gamma^{\hat r\hat s\hat t\hat u\hat v\hat w}  
F_{\hat t\hat u\hat v\hat w} + 24 \,\Gamma_{\hat t\hat u}  
F^{\hat r\hat s\hat t\hat u} ] \, {\rm d}\theta  
+ {\cal O}(\theta^4) \, , 
\end{eqnarray}
and agree completely with those obtained through gauge completion 
in \cite{backgr} and, for the spin-connection field, in \cite{CF}. 

\newpage
\setcounter{equation}{0}
\section{The four-form super field strength}
\noindent

The Wess-Zumino-Witten part of the action can be constructed by 
considering the most general ansatz for a four-form invariant 
under tangent-space transformations. Using the lowest-order 
expansions of the vielbeine 
(\ref{vielbeinexp}) and comparing with \cite{backgr} shows that 
only two terms can be present. Their relative coefficient is fixed 
by requiring that the four-form is closed, something that can be 
verified by making use of the Maurer-Cartan equations 
(\ref{maurercartan}). The result takes the form 
\begin{equation}
\label{wzwfourform}
F_{(4)} = \ft1{24} \Big[ 
E^{\hat r}\wedge E^{\hat s} \wedge E^{\hat t} \wedge E^{\hat u} 
F_{\hat r\hat s\hat t\hat u} 
- 12 \, \bar E \wedge \Gamma_{\hat r\hat s} E \wedge E^{\hat r} 
\wedge E^{\hat s} \Big]\, .
\end{equation}
To establish this result we also needed the well-known 
quartic-spinor identity in 11 dimensions. The overall factor in 
\eqn{wzwfourform} is fixed by comparing to the normalization of 
the results given in \cite{backgr}.

Because $F_{(4)}$ is closed, it can be written locally  as  
$F_{(4)}= {\rm d} B$. To low orders in $\theta$, the three-form 
$B$ is given by \cite{backgr}
\be
B= \ft16\, e^{\hat r}\wedge e^{\hat s}\wedge e^{\hat t} \,C_{\hat r 
\hat s \hat t} 
-\ft12 \, e^{\hat r}\wedge e^{\hat s}\wedge \bar \theta\,
\Gamma_{\hat r\hat s} D\theta  + {\cal O}(\theta^4) \, .
\ee
The general solution for $B$ can be found by again exploiting the 
one-forms with rescaled $\theta$ coordinates according to $\theta\to t\,
\theta$. Using \eqn{diffeq} we then find
\be 
\frac{\rm d}{{\rm d}t} F_{(4)} = - {\rm d}\Big( \bar\theta \,
\Gamma_{\hat r\hat s} E \wedge E^{\hat r}\wedge E^{\hat s}\Big) 
\,,
\ee
where again we made use of the quartic-spinor identity.  
This equation can directly be integrated so that we find for the 
three-form,
\be
B= \ft16\, e^{\hat r}\wedge e^{\hat s}\wedge e^{\hat t} \,C_{\hat 
r \hat s \hat t} 
-\int_0^1{\rm d}t\;  \bar\theta \,
\Gamma_{\hat r\hat s} E \wedge E^{\hat r}\wedge E^{\hat s}\, ,
\ee
which agrees to order $\theta^2$ with the result above. 
Furthermore the flat-space result (obtained by setting $F_{\hat r 
\hat s \hat t\hat u}= \omega^{\hat r\hat s}=0$),  
\be
B= \ft16\, e^{\hat r}\wedge e^{\hat s}\wedge e^{\hat t} \,C_{\hat r 
\hat s \hat t} 
-\ft12 \,\Big[  e^{\hat r}\wedge e^{\hat s}   + e^{\hat r} \wedge 
\bar\theta\,\Gamma^{\hat s}{\rm d}\theta + \ft13 \bar\theta\,
\Gamma^{\hat r}{\rm d}\theta \wedge \bar\theta\,\Gamma^{\hat 
s}{\rm d}\theta\Big] \wedge  
\bar \theta\, \Gamma_{\hat r\hat s} {\rm d}\theta       \,,
\ee
is correctly reproduced.

The supermembrane action is then written in terms of
the superspace embedding coordinates 
$Z^M(\zeta)=(X^{\hat\mu}(\zeta),\theta^{\hat a}(\zeta))$, which 
are functions of the world-volume coordinates $\zeta^i$ ($i=0,1,2$).
To all orders in anticommuting coordinates in an $AdS_4\times S^7$ 
or $AdS_7\times S^4$ background it is thus
given by
\be
S= -\int d^3\zeta \;\sqrt{-\mbox{det } g_{ij}(Z(\zeta))}
+\int_{M_3} B \, ,
\ee
where the induced worldvolume metric equals $g_{ij}=\Pi^{\hat r}_i 
\Pi^{\hat s}_j \, \delta_{\hat r\hat s}$ and 
$\Pi^{\hat r}_i=\partial Z^M/\partial\zeta^i\, E^{\hat r}_M$ is
the pullback of the supervielbein to the membrane worldvolume. This
action is invariant under local fermionic $\kappa$ transformations
\cite{BST} as well as under the superspace isometries 
corresponding to $osp(8\vert 4)$ or $osp(6,2\vert 4)$.

We have already emphasized that the choice of the coset representative 
amounts to adopting a certain gauge choice in superspace. The choice that 
we made in this paper connects directly to the generic 
11-dimensional superspace results, written in a Wess-Zumino-type 
gauge, in which there is no distinction between spinorial 
world and tangent-space indices. In specific backgrounds, such as 
the ones discussed here, there are gauge choices possible which 
allow further simplifications. For instance, one could 
reparametrize the anticommuting coordinate $\theta$ with an 
$x$-dependent matrix,
$\theta = \ell^{-1}(x) \,\theta^\prime$, where $\ell(x)$ is the coset 
representative of the bosonic part of the coset space introduced 
earlier, but written in the spinor representation. In that case 
$\ell(x)$ satisfies
\be
{\rm d} \ell^{-1} -\ft14 \omega^{\hat r\hat s} \G_{\hat r \hat s} 
\ell^{-1}  = - e^{\hat r}\, T_{\hat r}{}^{\!\hat s\hat t\hat u\hat v} 
F_{\hat s\hat t \hat u\hat v}  \ell^{-1}\,.
\ee
This equation can only be solved in a fully supersymmetric background. 
In this gauge (called the Killing gauge in \cite{KRR}) the 
expression for $D\theta$ is replaced by ${\rm d} 
\theta^\prime$, while ${\cal M}^2$ is appropriately redefined.
This choice, combined with a suitable
fixing of $\kappa$-symmetry, was reported
to simplify the corresponding action for the superstring
considerably \cite{kallosh}. Observe that, in this gauge,  
the 32 fermionic 
Killing vectors are given by $\partial/\partial\theta$. 

\setcounter{equation}{0}
\section{Discussion}
\noindent

In this paper we have constructed the superspace vielbein 
and three-form tensor gauge field for the $AdS_4\times S^7$ 
and $AdS_7\times S^4$ solutions of 11-dimensional supergravity 
to all orders in anticommuting coordinates. Our results provide
a strong independent check of the low-order $\theta$ results
obtained previously by gauge completion for general
backgrounds \cite{backgr,CF}. 
A great amount of clarity was gained by expressing our results 
in 11-dimensional language, where we were able to cover both
the $AdS_4\times S^7$ and the $AdS_7\times S^4$ solution in one
go. Expressed in terms of the  on-shell supergravity
component fields of the elfbein and three-form tensor gauge field, our
findings are in reassuring agreement with the gauge completion 
results obtained for general
backgrounds. Note that in the particular background we consider here, the
gravitino vanishes. We have no reasons to expect that the 
11-dimensional form of our results will coincide with the 
expressions for a generic 11-dimensional superspace (with the 
gravitino set to zero) at arbitrary orders in $\theta$.

The obtained superspace geometry was then utilized to construct the complete
M-theory two-brane action in $AdS_4\times S^7$ and $AdS_7\times S^4$
to all orders in $\theta$. This represents a further step
in the program of finding the complete anti-de Sitter background actions 
for the superstring \cite{MT,KRR} and the M2, D3 \cite{MT2} and M5-branes 
initiated for the bosonic part in \cite{KvPTK}. Certainly our
results for the $AdS_4\times S^7$ and $AdS_7\times S^4$ 
superspace geometries will be of use to construct the still missing 
M5-brane action in these backgrounds as well.

Moreover our results might prove to be a starting point for the construction
of a matrix model description of M-theory in $AdS_4\times S^7$ and 
$AdS_7\times S^4$ along the lines of \cite{DHN}, an option that 
was already discussed in \cite{backgr}.

\vspace{1cm}
\noindent {\bf Acknowledgements}: We thank 
R. Kallosh and J. Rahmfeld
for useful discussions. BdW and AS were supported in part 
by the European Commission TMR programme ERBFMRX-CT96-0045 in 
which AS is associated to K.U. Leuven.



\end{document}